\documentclass{article}
\PassOptionsToPackage{numbers, compress}{natbib}

\usepackage[preprint]{neurips_2021}

\usepackage[utf8]{inputenc} 
\usepackage[T1]{fontenc}    
\usepackage{hyperref}       
\usepackage{url}            
\usepackage{booktabs}       
\usepackage{amsfonts}       
\usepackage{nicefrac}       
\usepackage{microtype}      
\usepackage{xcolor}         
\usepackage{xspace}
\usepackage{amsmath}
\usepackage{graphicx}
\usepackage{epstopdf}

\usepackage{multirow}
\usepackage{soul}
\usepackage{subcaption}
\usepackage{booktabs}
\usepackage{widetable}
\usepackage{adjustbox}
\usepackage{makecell}
\usepackage{color, colortbl}
\usepackage{bbm}
\usepackage{times}
\usepackage{mathptmx}
\usepackage{algorithm}
\usepackage{algorithmic}

\title{A hybrid deep learning framework for Covid-19 detection via 3D Chest CT Images}

\author{
  Shuang Liang
 \\[.6ex]
  University of Science and Technology Beijing \\
  {\tt\small liangshuang@xs.ustb.edu.com}
}

\makeatletter
\DeclareRobustCommand\onedot{\futurelet\@let@token\@onedot}
\def\@onedot{\ifx\@let@token.\else.\null\fi\xspace}

\makeatother

\begin{document}

\maketitle

\begin{abstract}
    In this paper, we present a hybrid deep learning framework named CTNet which combines convolutional neural network and transformer together for the detection of COVID-19 via 3D chest CT images. It consists of a CNN feature extractor module with SE attention to extract sufficient features from CT scans, together with a transformer model to model the discriminative features  of the 3D CT scans. Compared to previous works, CTNet provides an effective and efficient method to perform COVID-19 diagnosis via 3D CT scans with data resampling strategy. Advanced results on a large and public benchmarks, COV19-CT-DB database was achieved by the proposed CTNet, over the state-of-the-art baseline approach proposed together with the dataset.
\end{abstract}

\section{Introduction}
\label{introduction}

The Coronavirus Disease 2019 (COVID-19), a highly infectious disease, has become a global pademic and posed serious threats to human worldwide. In order to prevent further spreading of COVID-19 and treat the infected patients instantly, various examination methods have been proposed for the diagnosis of COVID-19 \cite{alizadehsani2021risk,khadidos2020analysis}, of which rRT-PCR is usually considered as the golden standard for the diagnosis of COVID-19 \cite{long2020diagnosis}. However, due to the limitation of sample collection and transportation, the senstivity of rRT-PCR might encounter some problems. As reported by the World Health Organization (WHO), the lung infection was detected in the autosies \cite{sohrabi2020world}. Thus, medical imaging of chest radiography is proved to be useful for rapid COVID-19 detection \cite{xiong2020clinical}. Computed Tomography (CT) was considered as the precise examination tools since it provided a 3D view of organs especially the lung and could be used to locate the lesion areas \cite{dai2020ct}. In the CT examination, large volume of CT scans are obtained from persons suspected of COVID-19, which poses heavy workload on physicians and radiologists to diagnose COVID-19. There is a great need to develop some auxilary reliable methods in the medical imaging analysis of COVID-19. As demonstrated ub some studies, machine and deep learning methods might be a potential approach for the detection of COVID-19 \cite{song2021deep}.

However, there are two main problems in their works. First, the size of dataset used in these works is limited which makes it hard to avoid the overfitting problem and reduces the usability of their works. Second, the preprocess methods of 3D CT scans are not ideal. Some works only take the slices of CT scans that are suspected to have related sympotoms into account \cite{shah2021diagnosis,ko2020covid,afshar2021covid,wu2021covid}. But in actual use case, this method require physicians and radiologists to first screen these slices which might not be applicable. The other works take full volumn of CT scans into consideration. This method ensures that the discriminative features were included in the input data but make it hard to train models since the batch size could only be set to the single digits which might cause the unstable training process.

In this paper, we proposed a hybrid deep learning framework named CTNet that combines CNN and transformer together for the detection of COVID-19 via 3D chest CT images. Based on the hypothese that the relevant lesion area is continuous in lung and the features of lesion area is discriminative for the detection of COVID-19, we proposed a data resampling method to reduce the size of input data while preserving useful information. The contribution of our paper could be summarized as following:
\begin{enumerate}
\item A effective data resampling method was proposed for 3D CT scans to reduce the size of input while preserving sufficent information.
\item A hybrid deep learning framework is proposed for the detetion of COVID-19 which combines a CNN to mining effective features and a Transformer to conduct feature aggregation.
\item Good performances were achieved by the proposed framework with a macro F1 score of 88.24\% on the validation set of COV19-CT-DB dataset, which is 16 percent lead of that of the baseline method proposed along with the dataset. 
\end{enumerate}

The reset of this paper mainly focuses on the introduction of the proposed method and the corresponding results and discussion.

\section{Implemetation details}

\begin{figure*}[h]
    \begin{center}
        \includegraphics[scale=0.6]{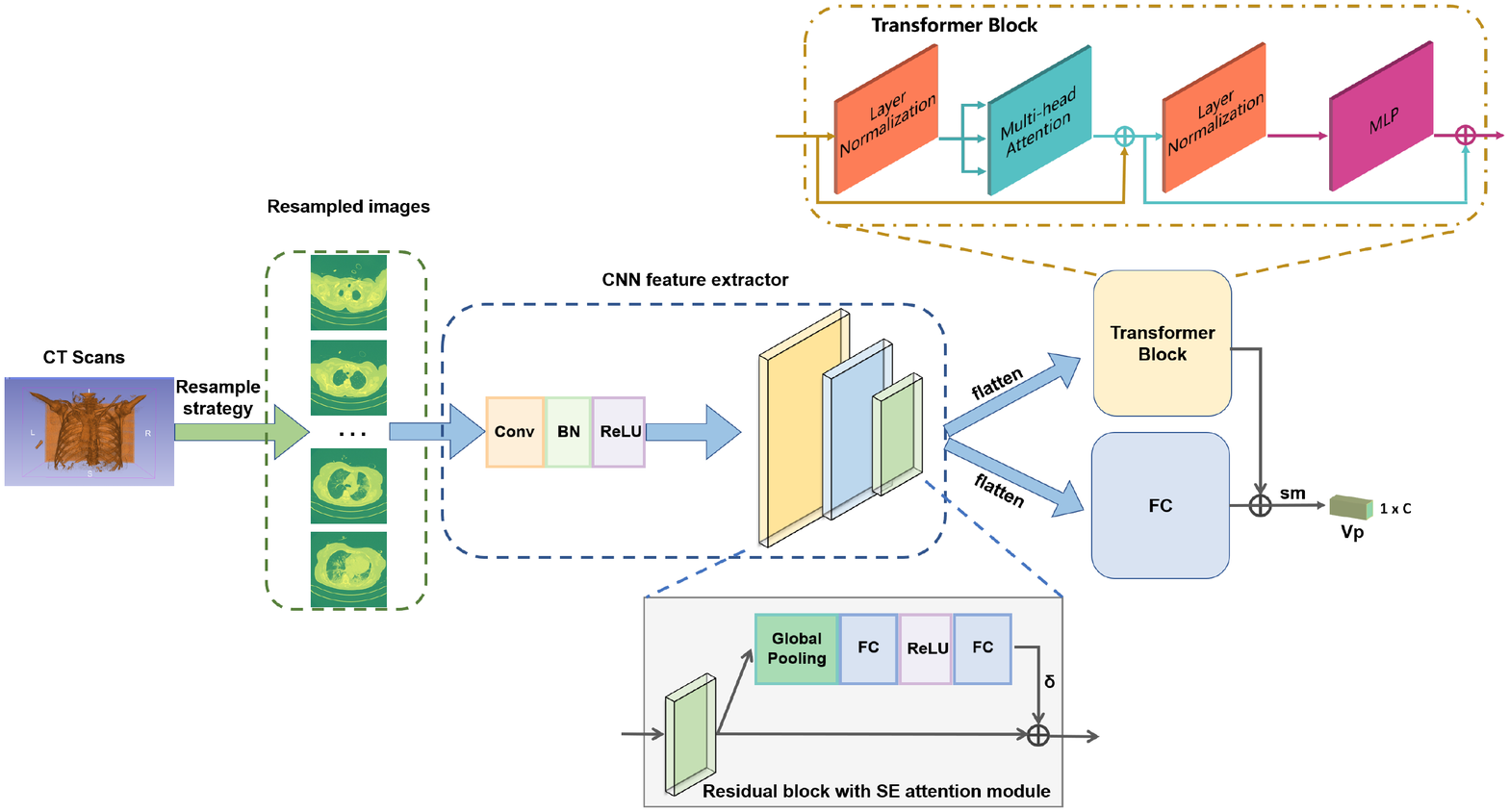}
    \end{center}
    \vspace{-2.5mm}
    \caption{The framework of the proposed CTNet.}
    \vspace{-0.5mm}
    \vspace{-2mm}
\end{figure*}

The proposed DL framework, called CTNet, is a hybrid DL method consisting of three componets: the input module with data resampling strategy, the CNN feature extractor module with SE attention module \cite{hu2018squeeze}, and the information aggregation module with the transformer and fully connected (FC) layer.
As shown in Figure 1, the 3D CT scans were resampled to a fixed number of CT slices as the input data, following is the CNN feature extractor that extracted sufficient and effective features. The extracted features were then flatten to vectors and sent to two branches, transformer brach and FC branch, and the vectors generated by the two branches were fused using element-wise-add (EWA) operation. The vector is then sent to the softmax function to predict the probability vector for the detection of COVID-19. 

\subsection{Data resampling strategy}
As we introduced in Introduction, the preprocess method of 3D CT scans matters a lot considering the actual application in clinical. Unlike the previous works that only take the annotated slices or take all the volumns of CT scans, we proposed a data resampling strategy that following the uniform distribution. Since the total number of CT scans varies from different cases, we firstly set a variable named totalnum to deal with this situation. The resampled number of CT scans is controled by the variable renumCT which could be adjusted according to the actual use case (especially the computation power and memory size). The algorithm contains two branches, one for resampling and the other for oversampling which is switchable according to the total number of CT scans and the resampled number of CT scans. The output is the indexes array of the 3D CT images. The process is summarized as following:

\begin{algorithm}
  \caption{Data resampled strategy for 3D CT scans}
  \label{alg1}
  \begin{algorithmic}
  \REQUIRE {$\rm{totalnum}, \rm{n}$}
  \ENSURE {$\rm{indexes}$(The indexes array of the resampled CT scans)}
  \STATE $\rm{renumCT} \gets \rm{n}$
  \STATE {$\rm{i} \gets 0$}
  \STATE {$\rm{indexes} \gets [ ]$}
  \IF{$\rm{totalnum} >= \rm{renumCT}$}
  \STATE $\rm{diff}=\rm{totalnum}/(\rm{renumCT}+1)$
  \WHILE{$i<renumCT$}
  \STATE $\rm{indexes}.append(i*\rm{diff})$
  \STATE $i+=1$
  \ENDWHILE
  \ELSE
  \WHILE{$i<renumCT$}
  \STATE $\rm{indexes}.append(random.choice(\rm{totalnum}))$
  \ENDWHILE
  \ENDIF
  \end{algorithmic}
\end{algorithm}

\subsection{CNN feature extractor with SE attention module}
The CNN feature extractor is a multi-stage residual neural network that incorporate the SE attention module to increase the discriminative ability. As shown in Figure. 1, the CNN can be seperated to two parts, the input convolution layer and the residual block. The number of the input channels for the input convolution layer is the same as the number of the resampled CT images. The following part is the residual block that adopted the SE module. The structure of the SE module is also demonstrated in Fig.1, which consists of a global pooling, two FC layers, the ReLU function, and the sigmoid function to generate attention vectors with the value range from 0 to 1. Given the resampled CT images as input, the extractor generates feature maps of N channels. Global pooling and flatten operations are performed to transform the feature maps to a feature vector.

\subsection{Information aggregation module}
As introduced previously, the 3D CT scans were resampled and sent to the CNN feature extractor. The output feature vector is then sent to two braches, the transformer block and the FC block, to fuse information and generate the final prediction. The structure of the transformer is shown in Fig.1, which consists of two layer normalization, a multi-head attention module and the multilayer perception module. The probability vectors generated by the two branches are fused using EWA operation to predict the diagnosis result.

\subsection{Training details}
The framework is trained in an end-to-end and from-scratch manner on a local machine with two 1080ti GPUs. The learning rate is set to 0.01, and is decreased following the step adjust strategy. The total epoches is set to 120, and the step epoches are 50 and 100. The number of resampled CT scans is set to 32 in our experiments and the batch size is set to 32. The size of each CT scan is resized to 224 $\times$ 224.

\section{Results}
\subsection{COV19-CT-DB database}
The dataset used in this paper for training and validation is the COV19-CT-DB dataset \cite{kollias2021mia,kollias2020deep,kollias2020transparent,kollias2018deep}, which consists of 5,000 chest CT scans that are annotated as COVID-19. Data was collected in the period from September 1, 2020 to March 31, 2021. The data were aggregated from different hospitals, containing anonymized human lung CT scans with signs of COVID-19 and without signs of COVID-19. The database was seperated into three set, the training set, the validation set and the test set, of which the training and validation set are publicly avaliable currently. The training set consists of 1560 3D CT scans, including 690 COVID-19 cases and 870 Non-COVID-19 cases. The validation set contains 374 3D CT scans, including 165  COVID-19 cases and 209 Non-COVID-19 cases. Each CT scan owns different numbers of CT scans, ranging from 50 to 700. The details were shown in Table. 1.

\begin{table}[H]
\caption{Distribution of the COV19-CT-DB database.}\label{tbl1}
\begin{center}
\begin{tabular}{cccccc}
\toprule
{KACD} &
{train} &
{val} &
{test} &{total}\\
\midrule
COVID-19 &	690 &	870	& - & 1560+\\
Non-COVID-19 &	165 &	209	& - & 374+\\
Total&855&1079&3455&5389 \\
\bottomrule
\end{tabular}
\end{center}
\end{table}

\subsection{Evaluation Metrics and experiments}
As introduced in the database, the 'macro' F1 score is used as the evaluation metric and the score of 0.70 is reported as the baseline score. The 'macro' F1 score is defined as the unweighted average of the class-wise/label-wise F1-scores.
In our experiments, we conduct two experiments to validate the proposed framework, one only uses the probability vector generated by the FC layer and the other uses the fused probability vector to obtain the diagnosis results.
The performance on the validation set can be seen from Table.2. Good performance was achieved by the CTNet, with a best 'macro' F1-score of $88.23\%$, which is $18.23$ percent higher than that of the baseline approach. Notably, the number of resampled CT scans is set to 32 in our experiments while that for the baseline is 700. Besides, the inference process of the CTNet cost only 380ms in diagnosing each case which makes it more applicable in actual use cases.

\begin{table}[H]
\caption{Performance of the CTNet framework}\label{tbl1}
\begin{center}
\begin{tabular}{ccc}
\toprule
{Method} &
{'macro' F1 Score} \\
\midrule
Baseline&70.00\% \\
CTNet (FC) &	87.96\% \\
CTNet (Fused) &	88.23\% \\
\bottomrule
\end{tabular}
\end{center}
\end{table}

\section{Conclusion}
We present a hybrid deep learning framework called CTNet for the detection of COVID-19 via 3D CT scans. The proposed CTNet provides a novel resampling strategy and network architecture for the effective mining of 3D CT scans. Experimental results demonstrated that our resampling strategy and framework achieved a good trade-off between speed and accuracy. Ablation studies showed CTNet's superior ability to deal with COVID-19 detection with less requirement of computation power and data.

\bibliographystyle{plainnat}
\bibliography{main.bib}

\end{document}